\documentclass{v16nufact}

\def\Journal#1#2#3#4{{#1} {\bf #2}, #3 (#4)}

\def\NIMA{{\em Nucl. Instrum. Methods} A}

\def\PRD{{\em Phys. Rev.} D}

\def\be{\begin{equation}}
\def\ee{\end{equation}}
\def\bea{\begin{eqnarray}}
\def\eea{\end{eqnarray}}

\newcommand{\Photo}{}

\begin{document}
\vspace*{4cm}
\title{T2K NEAR DETECTOR CONSTRAINTS FOR OSCILLATION RESULTS}

\author{L. HAEGEL for the T2K collaboration}

\address{DPNC, \'Ecole de Physique, Quai Ernest Ansermet 24\\
1205 Gen\`eve, Switzerland}

\maketitle
\abstracts{The T2K experiment is a long-baseline accelerator neutrino experiment using a near detector complex ND280 and a far detector, Super-Kamiokande. Neutrino interactions are detected by Cherenkov light in Super-Kamiokande in order to measure neutrino oscillation parameters. The accuracy of the oscillation parameter measurements depends on our knowledge of neutrino interactions, the neutrino flux and the detector response. ND280 is composed of a tracker and several sub-detectors designed to characterise the neutrino beam before oscillation, which allow us to constrain the uncertainties on the neutrino interaction and the accelerator flux models. In this talk we present the result of a fit on the ND280 data, as well as a study of the robustness of the fit to the choice of neutrino interaction model. The constraint obtained on the oscillation parameters is also discussed.}

\section{The T2K experiment}

T2K~\cite{t2k} is a long baseline neutrino oscillation experiment located in Japan. 
Neutrinos are produced in the J-PARC accelerator complex in Tokai, Ibaraki, by sending protons on a graphite target.
The horn current polarity can be chosen to be negative or positive in order to create a neutrino or antineutrino-enhanced beam.
The beam is sent to the off-axis far detector Super-Kamiokande (Super-K) located 295 Km west in the Kamioka mine in Hida, Gifu. 
There, neutrino interactions are detected through Cherenkov effect in order to probe $\nu_{\mu}$ disappearance and $\nu_{e}$ appearance.
Oscillation measurements require accurate estimation of the number of events, therefore the flux and interactions mechanisms must be known precisely. 
The off-axis near detector ND280 is located 280 m from the J-PARC target in order to select non-oscillated neutrino events and constrain the flux and cross-section models.

\begin{figure}[h]
\centerline{\includegraphics[width=0.4\linewidth]{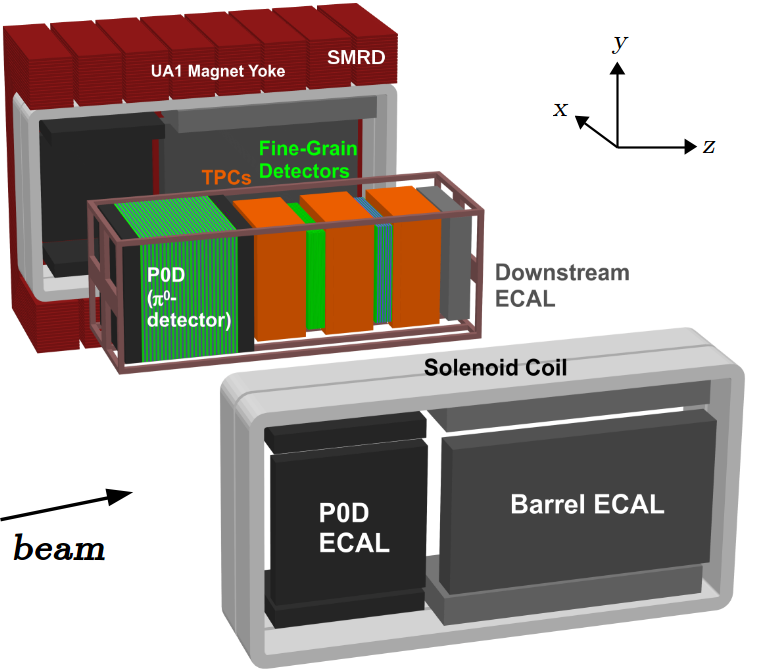}}
\caption[]{Exploded view of the off-axis near detector ND280.}
\label{fig:nd280}
\end{figure}

ND280 contains a tracker made of three Time Projection Chambers (TPCs) with two Fine Grain Detectors (FGDs) interlaid shown on Figure~\ref{fig:nd280}.
The first FGD is made of 15 modules, one module being made of a first layer with 192 plastic scintillator bars aligned in the X-direction, and a second layer with 192 bars aligned in the Y-direction.
The second FGD is made of 8 scintillator modules identical to FGD1, with 7 water modules placed between every scintillator modules.
In addition to the tracker, the $\pi^0$ detector (P0D) is made of modules of scintillator and water interlaid, with copper and leads sheets to contain the electromagnetic shower created by the $\pi^0$.
Around those detectors are placed several electromagnetic calorimeters, namely the P0D ECal around the P0D, the Barrel ECal around the tracker and the Downstream ECal after the last TPC.
The tracker, P0D and ECals are located inside the basket of the UA1 magnet, providing a magnetic field to distinguish the electric charge of particles and measure their momentum. 
The Side Muon Range Detector is interlaid with the UA1 magnet plates and provides tagging for particles coming from outside ND280.
In addition, the on-axis near detector INGRID is placed under ND280 and monitors the beam direction in real time.

\section{Oscillation measurements}
\label{sec:osc}

\begin{figure}[h!]
\begin{minipage}{0.4\linewidth}
\centerline{\includegraphics[width=1.\linewidth]{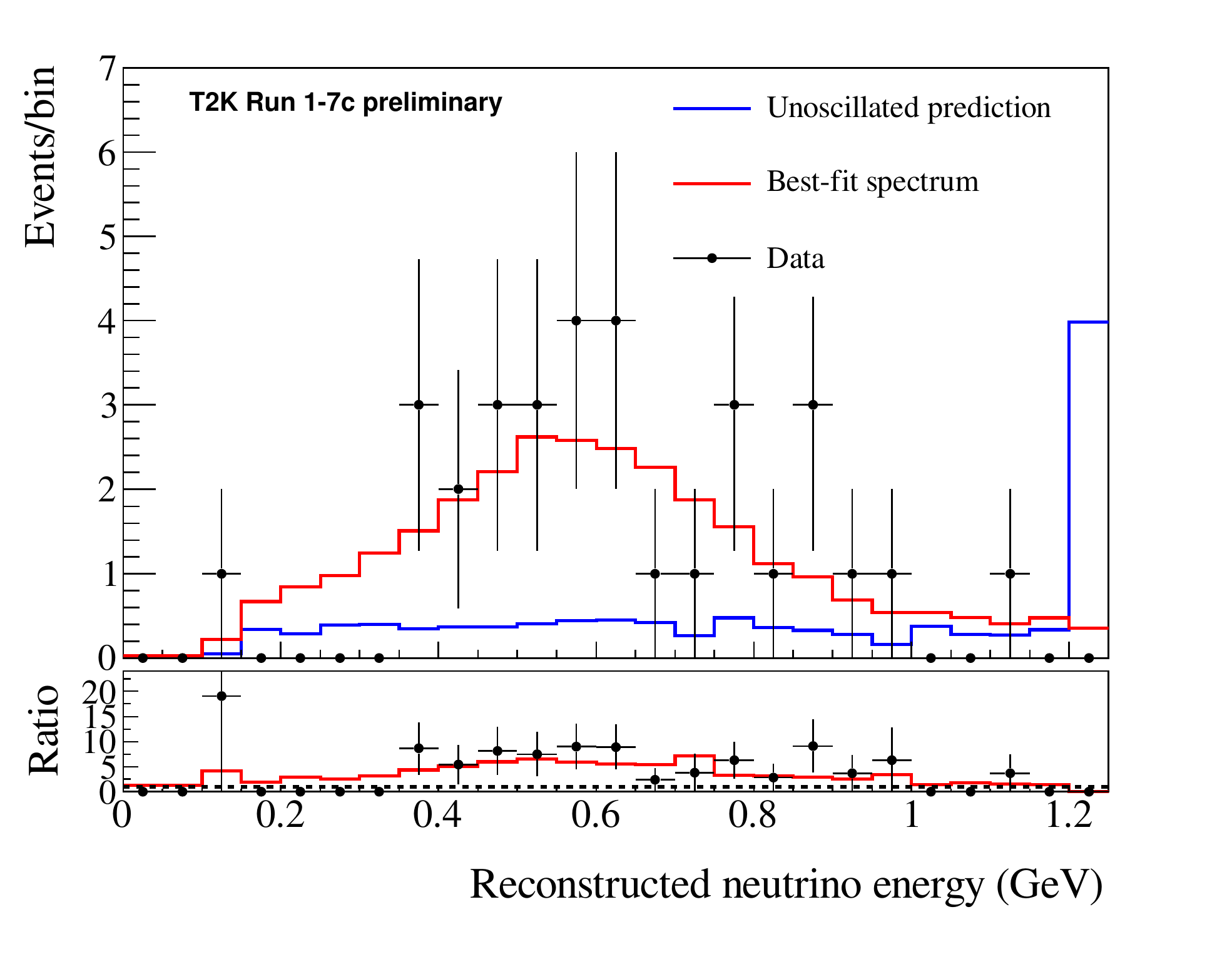}}
\end{minipage}
\hfill
\begin{minipage}{0.6\linewidth}
\centerline{\includegraphics[width=1.\linewidth]{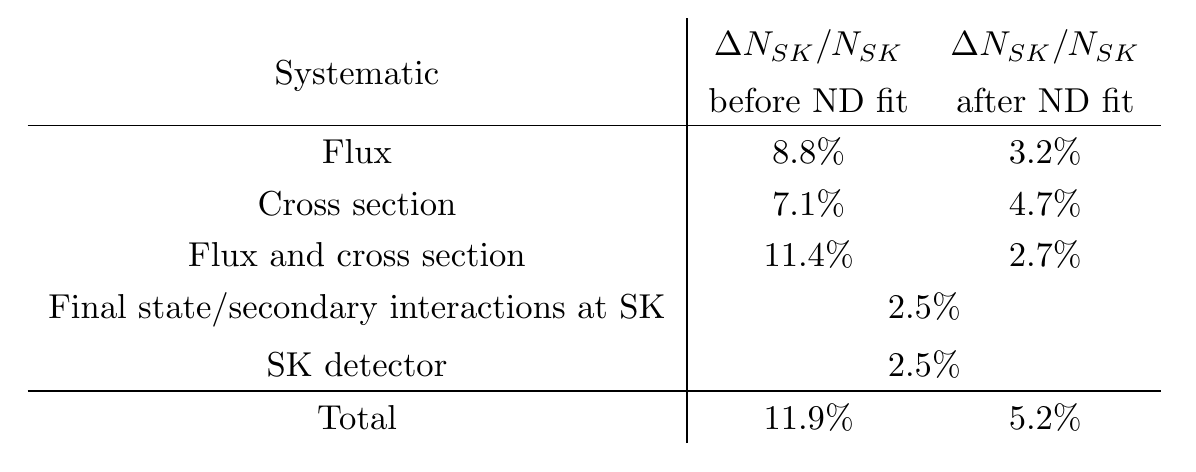}}
\end{minipage}
\caption[]{Left: spectra of $nu_e$ selected events, points are data, red is Monte-Carlo oscillated with best-fit prediction, blue is unoscillated Monte-Carlo. Right: uncertainties on the predicted $\nu_e$ event rate from each category. T2K run 1-7 preliminary.}
\label{fig:sknue}
\end{figure}

T2K aims at measuring four of the six parameters of the PMNS matrix~\cite{nujointfit}: $sin^2 \theta_{23}$, $sin^2 \theta_{13}$, $\Delta m_{23}^2$ and $\delta_{CP}$. 
The estimation of the parameters is done by fitting the expected number of selected Super-K charged-current interactions to the data.
Two kind of statistical methods are used: a semi-frequentist fit of Super-K events, and a Markov Chain Monte Carlo framework fitting ND280 and Super-K events.
Both uses the same binned likelihood in equation \ref{eq:llh}:
\begin{equation}
\small
	- ln (L) = \sum\limits_{i}^{N \ bins} \ N_i^p(\phi, \sigma, \epsilon) \ - \ N_i^d \ + \ N_i^d \ ln\left(\frac{N_i^d}{N_i^p(\phi, \sigma, \epsilon)}\right) 
+ \sum\limits_{i}^{\phi, \sigma, \epsilon \ pars} \ \sum\limits_{j}^{\phi, \sigma, \epsilon \ pars} \ \Delta (\phi, \sigma, \epsilon)_i \ (V_{\phi, \sigma, \epsilon}^{-1})_{i,j} \ \Delta (\phi, \sigma, \epsilon)_j 
\label{eq:llh}
\end{equation}
where $N_i^p$ is the expected number of events, $N_i^d$ is the real number of events, $V_{i,j}$ are the covariance matrices constraining the systematics parameters on the flux $\phi$, the cross-sections $\sigma$ and the detector efficiency $\epsilon$. \\
The expected numbers of events in each bin $i$ is: 
\begin{equation}
N_{SK}^{\nu_{\beta}}(i) = \phi_{SK}^{\nu_{\alpha}}(i) \ \sigma^{\nu_{\beta}}(i) \ \epsilon_{SK}^{\nu_{\beta}}(i) \ P_{PMNS} (\nu_{\alpha} \rightarrow \nu_{\beta})(i) 
\label{eq:nsk}
\end{equation}
A strong degeneracy can be seen between the flux, cross-section and detector efficiency on one hand, and the PMNS parameters of interest on the other hand. 
In order to constrain as accurately as possible the oscillation parameters, the models for each category of nuisance parameters must be known as precisely as possible.
The systematics uncertainties are strongly reduced after fitting a selection of data in ND280 as shown on Figure~\ref{fig:sknue} for the fit to run 1-7c data. 
It shows the spectra of selected neutrino beam mode $\nu_e$ events with the Monte-Carlo prediction in the oscillated and unoscillated case, as well as the table of the uncertainty on the total $\nu_e$ rate for each category of systematics before and after the ND280 fit. 
The uncertainties on other samples used in the oscillation fit (neutrino beam mode $\nu_{\mu}$ events, antineutrino beam mode $\nu_{\mu}$ and $\nu_e$ events) are of the same order of magnitude.

\section{Systematic uncertainties}
\label{sec:syst}

\subsection{Flux model}

\begin{figure}[h!]
\begin{minipage}{0.46\linewidth}
\centerline{\includegraphics[width=1.\linewidth]{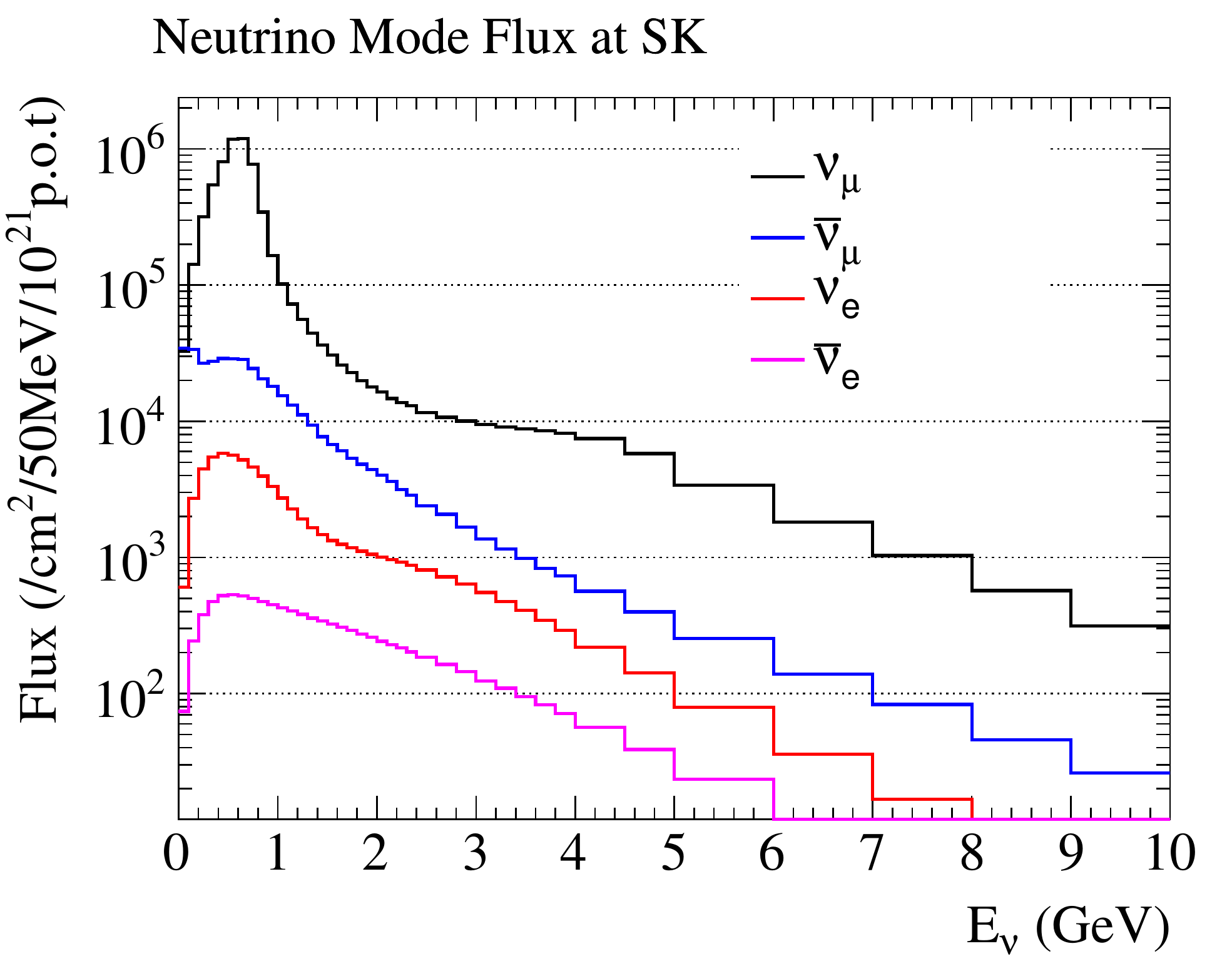}}
\end{minipage}
\hfill
\begin{minipage}{0.54\linewidth}
\centerline{\includegraphics[width=1.\linewidth]{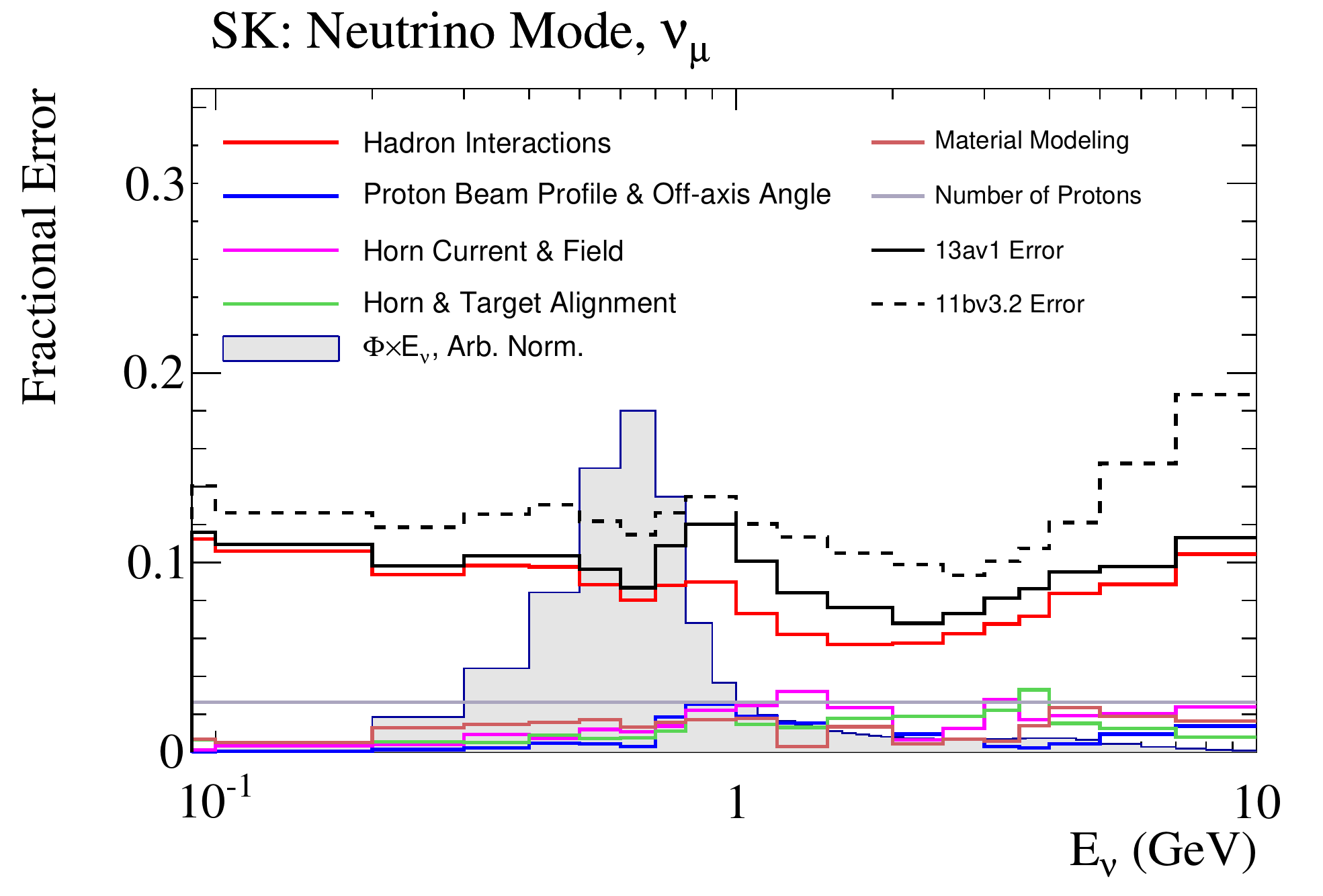}}
\end{minipage}
\caption[]{Left: flux spectrum prediction in Super-K in neutrino-enhanced beam mode. Right: uncertainties on the predicted $\nu_{\mu}$ flux at Super-K (13av1 is the latest version of T2K flux simulation used for the figure, 11bv3.2 is the previous version). T2K run 1-7 preliminary.}
\label{fig:flux}
\end{figure}

The interactions of protons on the graphite target are simulated with the FLUKA software tuned with NA61 data~\cite{na61}.
They create $\pi^{+/-}$ and $K^{+/-}$ focused through three horns selecting a neutrino or antineutrino beam according to their current polarity.
Hadrons are directed to the decay volume to create a (anti-)$\nu_{\mu}$ enhanced beam as shown on the predicted neutrino mode flux spectrum at Super-K on the left image of Figure~\ref{fig:flux}.
The propagation of the hadrons after the target is simulated with the GCALOR and Geant4 softwares, tuned with beam monitors data. \\
Six main sources of uncertainties have been found on the flux model obtained, amongst which two concern the proton beam: the profile of the beam and the number of protons sent on the target.
The hadron interactions in the target are another source of uncertainty, then two other arise from the horns: their alignment with the target and the current creating the magnetic field.
The latest uncertainty is on the modelling of the material surrounding.
The total uncertainty, before fitting ND280 data, is about 10\% at T2K flux peak as shown on the right image of Figure~\ref{fig:flux}. \\
The ND280 fit uses 50 true neutrino energy bins to constrain the flux model.
The binning has variable bin widths: it contains 11 bins for the (anti-)$\nu_{\mu}$ beam content in (anti-)neutrino beam mode, 5 bins for the wrong-sign (anti-)$\nu_{\mu}$ beam content in neutrino(anti-neutrino) beam mode, 7 bins for the (anti-)$\nu_{e}$ beam content in (anti-)neutrino beam mode, and 2 bins for the wrong-sign (anti-)$\nu_{e}$ beam content in neutrino(anti-neutrino) beam mode.

\subsection{Cross-section model}

T2K uses the NEUT software~\cite{neut} to model neutrino interactions, with a nuclear model chosen after fitting external data~\cite{externaldatafit}.
Version 5.3.2 models the distribution of charges in the nucleus as a global Relativistic Fermi Gas (RFG), where nucleons can fill the momentum states up to a constant Fermi level~\cite{xsec}.
The model also includes a medium polarisation due to long-range correlations created by particule-hole propagation in the nucleus, named relativistic Random Phase Approximation (RPA). 
This section describes the interaction modes implemented in NEUT and explicits the free parameters allowed to float in the ND280 fit. \\

\begin{figure}[h!]
\begin{minipage}{0.5\linewidth}
\centerline{\includegraphics[width=1.\linewidth]{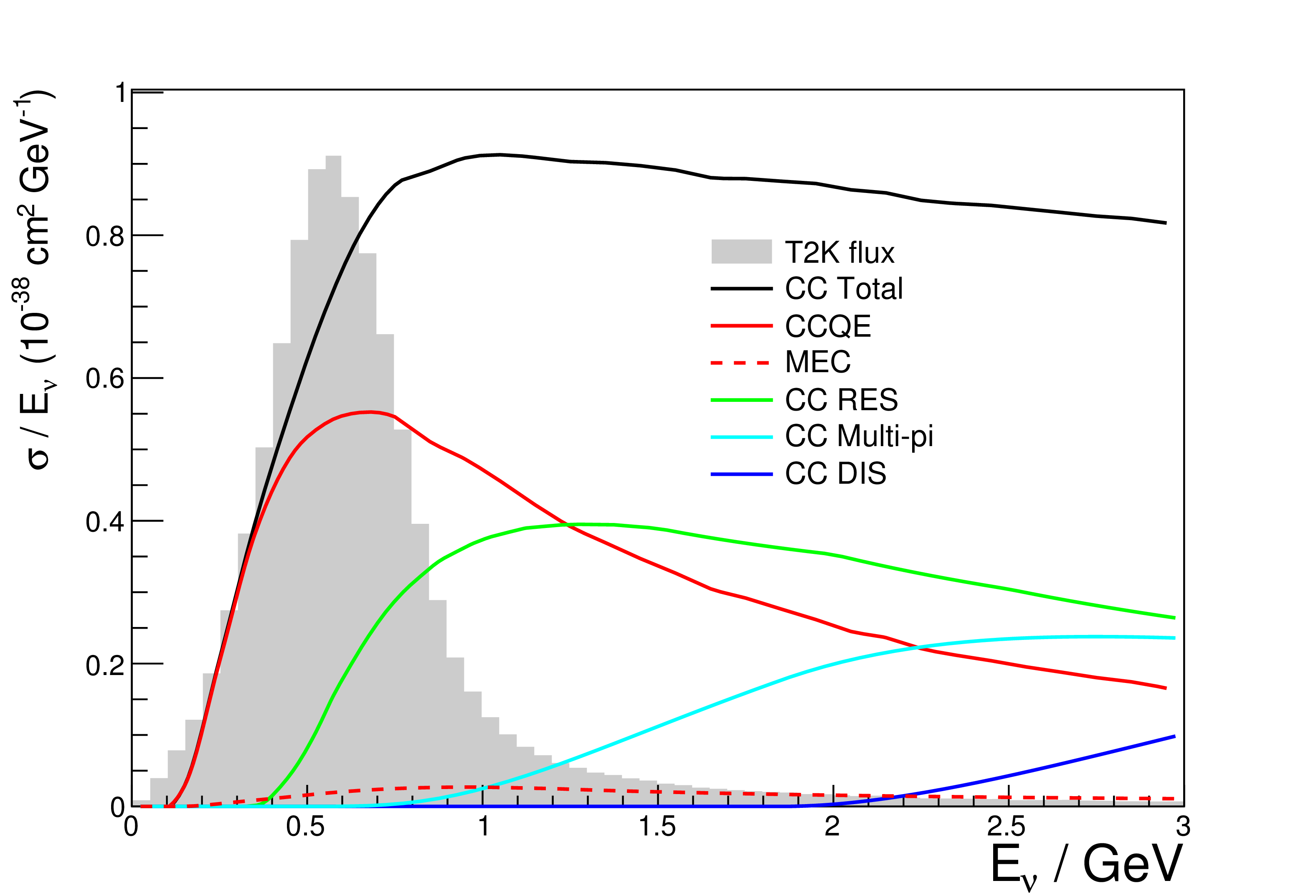}}
\end{minipage}
\hfill
\begin{minipage}{0.5\linewidth}
\centerline{\includegraphics[width=1.\linewidth]{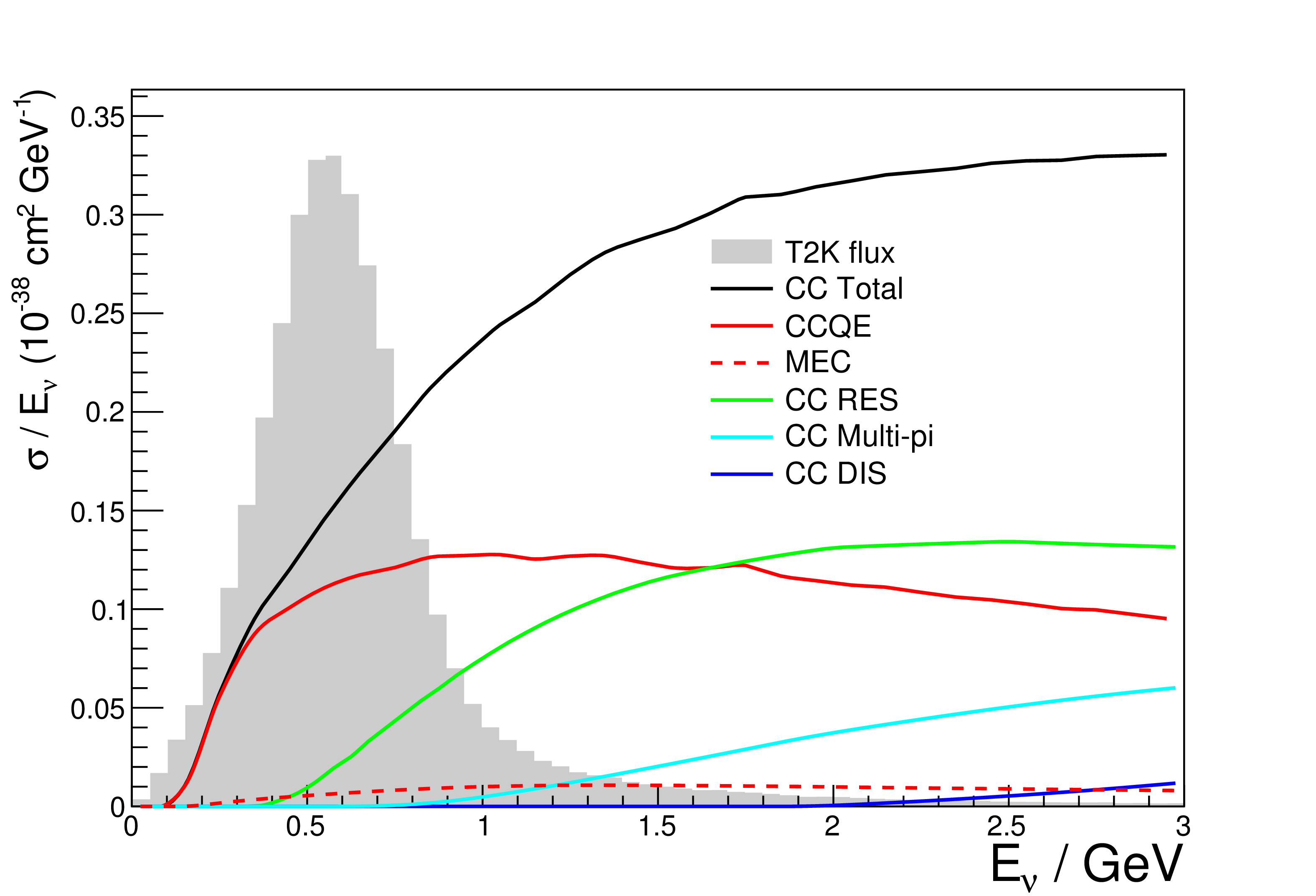}}
\end{minipage}
\caption[]{Neutrino charged-current cross-section interactions relevant at T2K energies. Left: $\nu_{\mu}$. Right: $\bar{\nu_{\mu}}$. T2K run 1-7 preliminary.}
\label{fig:xsec}
\end{figure}

The dominant process at T2K flux peak energy is Charged-Current Quasi-Elastic (CCQE) interactions as shown on Figure~\ref{fig:xsec}.
The CCQE model contains five parameters: an axial mass acting on the shape and the normalisation of the cross-section, two separate parameters for the Fermi level of carbon and oxygen atoms and two other parameters for the binding energy of carbon and oxygen.
Two particles - two holes (2p2h, or MEC for Meson-Exchanged Current on Figure~\ref{fig:xsec}) interactions occur when a neutrino interacts with a correlated pairs of protons, leading to the ejection of two protons.
As the protons are almost always under the Cerenkov threshold, Super-K is not able to detect them and 2p2h events appear CCQE-like and can induce a bias if the neutrino energy is computed according to this hypothesis.
They are included in NEUT with the Nieves model~\cite{2p2hn}, the free parameters being the normalisation of the 2p2h cross-section on carbon, the relative normalisation of the cross-section on water compared to carbon and the relative normalisation of the cross-section for $\nu$ compared to $\bar{\nu}$. \\
Charged-Current (CC) and Neutral-Current (NC) resonant pion production are implemented with the Rein-Sehgal model~\cite{xsec}, the free parameters being the resonant axial mass, the normalisation and shape parameter $C_A^5$ and the background from isopsin 1/2 interactions.
The CC coherent pion production cross-section normalisation is estimated with two parameters for carbon and oxygen respectively, while there is only one free parameter for the NC coherent pion production cross-section normalisation.
The normalisation of the cross-section for all other NC interactions is also evaluated in the ND280 fit.
The CC Deep Inelastic Scattering (DIS) and multi-pion production cross-section share a single normalisation parameter.
Finally, the relative $\nu_e / \nu_{\mu}$ cross-section normalisations are estimated with two parameters, one for $\nu$ and another one for $\bar{\nu}$.
Final state interactions such as pion absorption, scattering or production in the nucleus after the interaction occurs can lead to a misreconstruction of neutrino kinematic parameters.
They are implemented with a cascade model in NEUT, including six parameters constrained in the ND280 fit.

\section{ND280 fit}

\subsection{Fit results}

\begin{figure}[h]
\begin{minipage}{0.46\linewidth}
\centerline{\includegraphics[width=1.\linewidth]{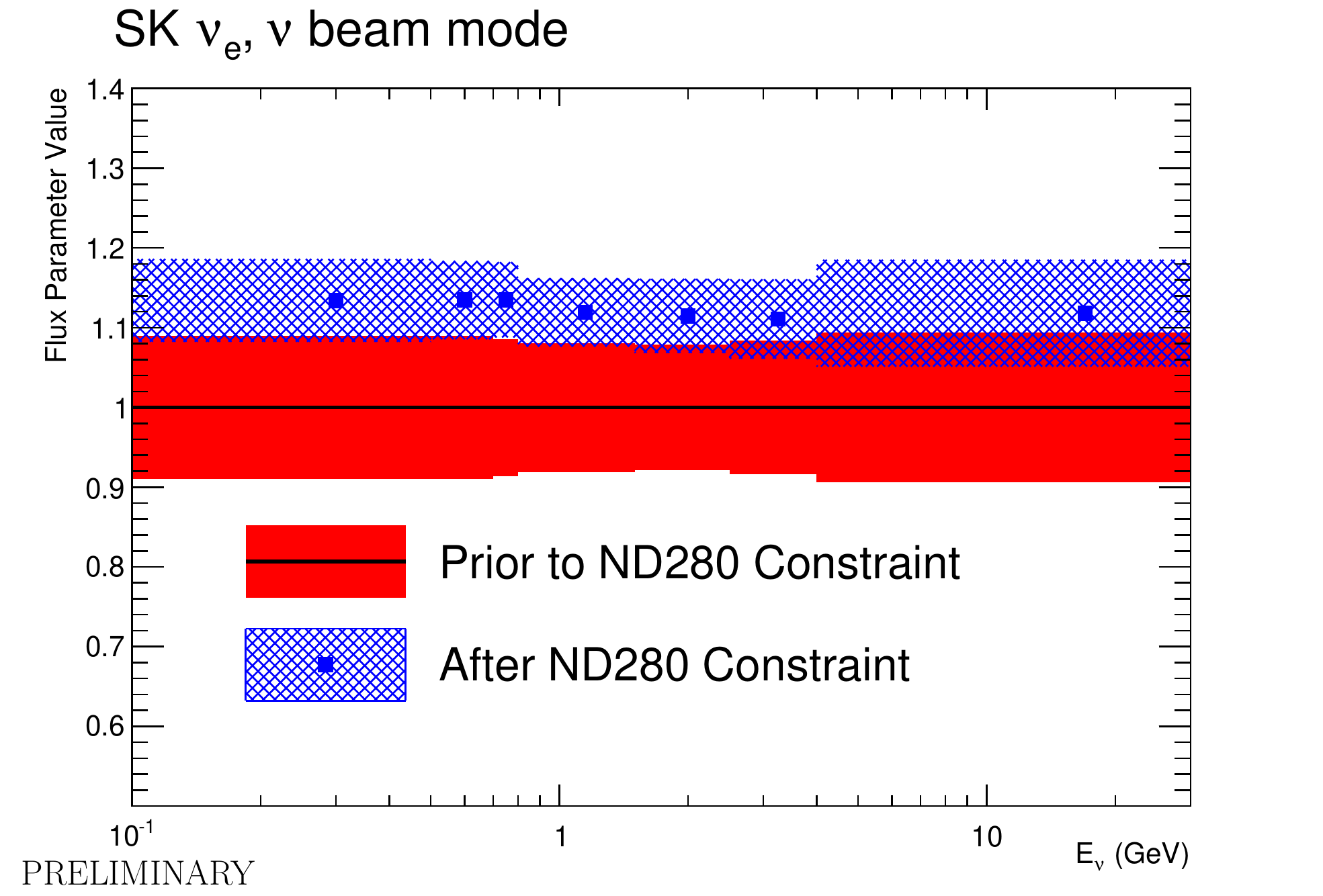}}
\end{minipage}
\hfill
\begin{minipage}{0.54\linewidth}
\centerline{\includegraphics[width=1.\linewidth]{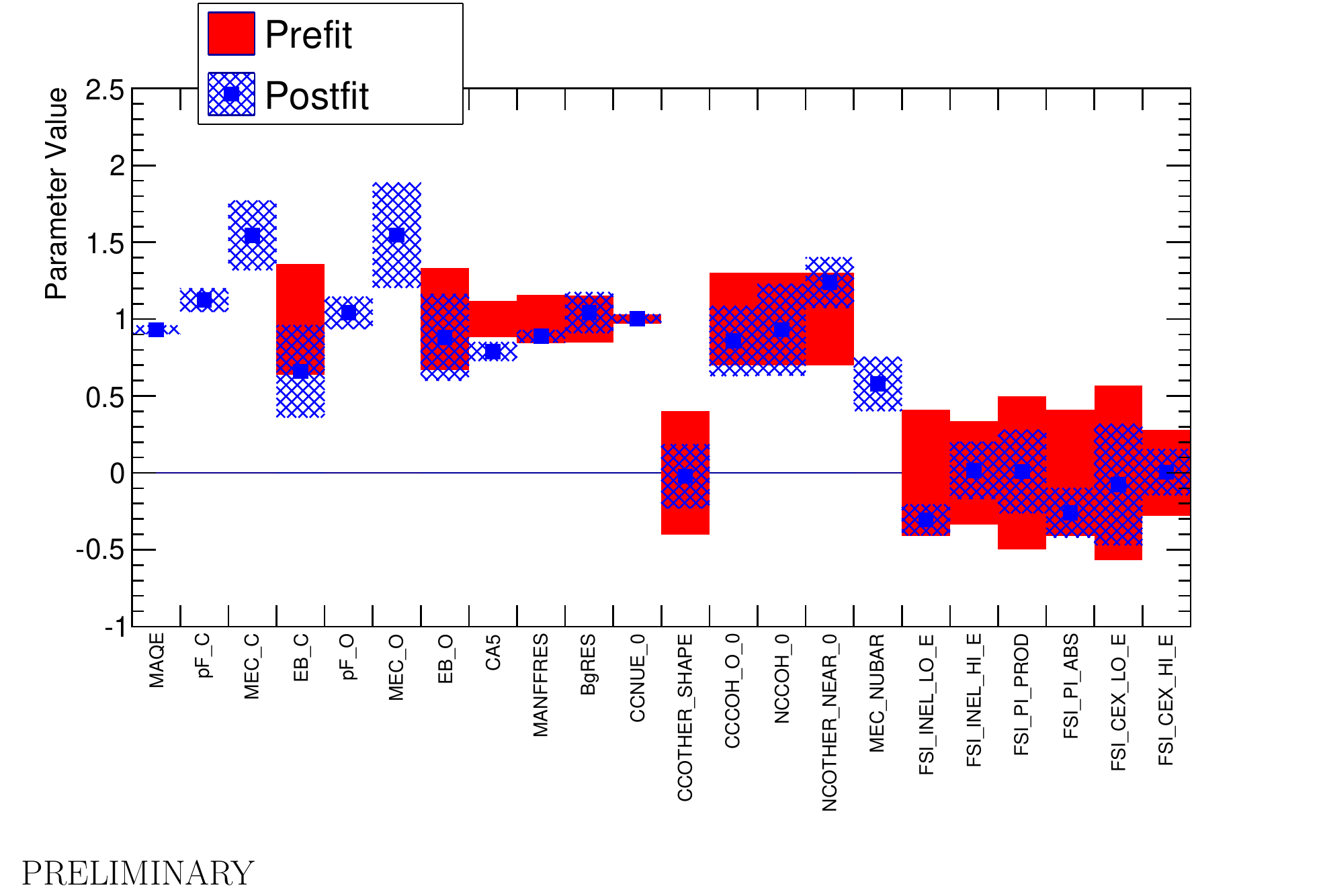}}
\end{minipage}
\caption[]{Comparison of the pre-ND280 fit and post-ND280 fit constraints on the parameters of the models described in Section~\ref{sec:syst}. Left: $\nu_e$ flux in neutrino mode at Super-K. Right: cross-section parameters. }
\label{fig:nd280fit}
\end{figure}

The selection of (anti-)$\nu_{\mu}$ CC interactions in ND280 requires the presence of a (anti-)muon-like track in a TPC that originates in the fiducial volume of the upstream FGD. 
The (anti-)muon track must be the particle of highest momentum, and there is a veto if tracks upstream the FGD are present.
This inclusive selection is done for $\nu_{\mu}$ events in neutrino beam mode, $\bar{\nu_{\mu}}$ events in antineutrino beam mode and wrong-sign $\nu_{\mu}$ events in antineutrino beam mode.
There are two selections is antineutrino beam mode as the wrong-sign neutrino contamination is higher than in neutrino beam mode, and must be evaluated as Super-K cannot distinguish the electric charge of particles.
The selections are afterwards separated in several categories according to their topology: in neutrino mode they are the $CC-0\pi$ category where only one muon track is seen, the  $CC-1\pi$ where one muon and one pion-like tracks are seen and the $CC-other$ category contains the other events.
For the antineutrino mode, the $\bar{\nu_{\mu}}$ and $\nu_{\mu}$ selections are broken into the $CC-1 \ track$ category where only one (anti-)muon track is seen, and the $CC-N \ tracks$ category contains the other events. \\
The fit of those selected Monte-Carlo events to the data minimise the same likelihood as the one given in Section~\ref{sec:osc} (without applying the oscillation probability to the events, as ND280 is too close to the target for oscillations to occur).
Figure~\ref{fig:nd280fit} shows the comparison of the pre-ND280 fit and post-ND280 fit constraints on the parameters of the models described in Section~\ref{sec:syst}.
The left image is the comparison for the $\nu_e$ flux in neutrino mode at Super-K, and the right image is the comparison for the interaction parameters.
The red band shows the prior Gaussian constrain on the parameters.
All flux parameters have a prior constraint while some cross-section parameters, e.g. the CCQE axial mass $M_A^{QE}$, the Fermi levels $p_F$ and the 2p2h normalisations $MEC$ have an unconstrained flat prior.
The blue dot is the best fit point obtained by the ND280 fit, while the blue band is the postfit uncertainty on the prior.
The flux parameters are pulled up by the fit, but their uncertainty decreases as the blue band is smaller than the red band (the same behavior is observed for the flux parameters for the $\nu_{\mu}$, $\bar{\nu_{\mu}}$ and $\bar{\nu_{\mu}}$ samples).
Most cross-section parameters best-fit points lie inside the prior uncertainty, and the uncertainty is always reduced by the fit.
Correlations between the parameters are not shown here, but the ND280 fit output is a covariance matrix including flux and cross-section parameters correlations used to constrain the oscillation fit at Super-K.
For example, the Super-K-only oscillation fit uses the post-ND280 fit covariance matrix for the $V_{i,j}$ elements of the likelihood formula on Equation~\ref{eq:llh}, and fits a Monte-Carlo tuned to the best fit points of the ND280 fit to lower the uncertainties on the nuisance parameters.

\subsection{Robustness of the fit to the cross-section model}

In order to evaluate if the ND280 fit is sensitive to the cross-section model used to generate the Monte-Carlo events, the fit has been performed with fake data simulating different cross-section models from the one described in Section~\ref{sec:syst}.
The Monte-Carlo was re-tuned in order to mimic the desired model for the fake data.
Different nuclear description have been tested: instead of the global RFG, a local RFG or Benhar's spectral function for the distribution of nucleons have been used.
Two alternatives to Nieves 2p2h model implementation have been tested: fitting Martini's 2p2h model~\cite{2p2hm} and adding a shape parameter for Nieves 2p2h cross-section.
The RPA polarisation has been replaced by an effective RPA where a function with several parameters was describing the model. \\
The interaction models were modified one at a time, and the fit to ND280 fake data was performed.
If a different constraint of the interaction parameter was found, the Super-K Monte-Carlo was tuned to the model tested, and the oscillation fit was redone in order to check if it was leading to a different estimation of the PMNS parameters.
It was found that the ND280 fit was giving different results every times the model was changed.
The fit to Super-K fake data revealed a bias in the estimated oscillation parameters when using the local RFG, which was incorporated into the post ND280 fit covariance matrix.
Other models were also leading to a different estimation of oscillation parameter, however the bias was very small with regards to statistical uncertainties. 
Therefore the uncertainty due to the choice of the interaction model is negligible for the moment, but will become important as the experiment runs.

\section{Summary and future prospects}

In order to achieve accurate estimation of oscillation parameters, the models used to estimate the number of predicted events must be known with a great precision.
T2K achieves this by fitting non-oscillated events in its off-axis near detector ND280 to constrain the flux and cross-section parameters.
The uncertainty on the predicted number of events at Super-K due to those sources of systematics decreases from 11.4\% to 2.7\% for the $\nu_e$ events in neutrino beam mode, with similar uncertainties for the other samples ( $\nu_{\mu}$ and $\nu_e$ events in neutrino and antineutrino beam modes). \\
The robustness of the fit to different interaction models has been studied.
It was found that the oscillation fit results were depending on the model, leading to include the uncertainties due to the local RFG model on the post-ND280 fit covariance matrix.
Although other models have a small effect with regards to statistical uncertainty, this study shows that the cross-section models will become a dominant uncertainty for future experiments. \\
In order to improve the constraint on the systematical uncertainties, new implementations are under investigation for the ND280 fit.
The selection currently used has a limited phase-space as it requires that the event starts in the FGD and continues in the downstream TPC.
A 360 degrees selection is being performed by using timing information between the detectors and ECal tagging.
New selections to add to the fit are also studied: CC (anti-)$\nu_e$ events in the tracker and CC (anti-)$\nu_{\mu}$ events in the P0D.
Improving the models themselves is another way to reduce our uncertainties, and a better tuning of the flux model with the NA61 long target results~\cite{na61l}, as well a better modelling of the neutrino interaction in NEUT are being implemented.

\end{document}